\documentstyle[aps,preprint,prb]{revtex}
\begin{document}
\def\vk{\vec k} 
\def\br{{\bf r}}
\title{\bf P-wave Pairing and Colossal Magnetoresistance \\
in Manganese Oxides}
\author{Yong-Jihn Kim$^{\dagger}$}
\address{Department of Physics,  Korea Advanced Institute of Science and 
Technology,\\
 Taejon 305-701, Korea}
\maketitle
\begin{abstract}
We point out that the existing experimental data of most manganese oxides
show the {\sl frustrated} p-wave superconducting condensation in the 
ferromagnetic phase in the sense that the superconducting coherence is not
long enough to cover the whole system.
The superconducting state is similar to the $A_{1}$ state in superfluid He-3. 
The sharp drop of resistivity, the steep jump of specific heat, and the
gap opening in tunneling are well understood in terms of the p-wave pairing. 
In addition, colossal magnetoresistance (CMR) is naturally explained
by the superconducting fluctuations with increasing magnetic fields.
The finite resistivity may be due to some magnetic inhomogeneities. 
This study leads to the possibility of room temperature superconductivity.  
\end{abstract} \vskip 4pc
\noindent

\vskip 2pc
$^{\dagger}$ Present address : Department of Physics, Bilkent University\\
Bilkent 06533, Ankara, Turkey
\vfill\eject
\section*{\bf 1. Introduction} 

Recently, much attention has been paid to the manganese oxides since  
the observation of colossal magnetoresistance (CMR).$^{1-4}$
The nearly half-metallic nature of ferromagnetic phase was also reported.$^{5-8}$
In addition, the layered manganite was found.$^{9}$
These systems are strongly correlated and show antiferromagnetic-ferromagnetic 
transition and metal-insulator transition simultaneously.
These phenomena were explained by the double exchange mechanism.$^{10-12}$ 
However, there is still no consensus as to the theoretical understanding  
because of the strong interaction between
spin, charge, and lattice degrees of freedom.$^{13,14}$

In this letter, we propose a {\sl frustrated} p-wave superconducting state for 
the ferromagnetic phase of manganese oxides. 
In fact, the existing experimental data seem to manifest the 
p-wave superconducting condensation.
In this scenario, colossal magnetoresistance (CMR) is naturally 
explained by the superconducting fluctuations with increasing magnetic fields.
The superconducting state is similar to the well-known 
$A_{1}$ state which is realized in the superfluid He-3.$^{15}$ 
Only majority spin up electrons are paired. 
Because of a strong Hund coupling between the $e_{g}$ conduction electrons and 
$S=3/2$ core spins formed by the $t_{2g}$ electrons, the ferromagnetism appears
and the conduction electrons are highly spin-polarized in the ground state.
If the majority spin up electrons are paired by the currently unknown 
mechanism, the energy of the ground state will be even lower. 
Presumably, both the electron-phonon interaction and the magnetic interaction
contribute significantly.  The gap parameter is very anisotropic
and has point nodes at the points of intersection of the Fermi surface with
the spontaneous anisotropy axis.
Whereas the minority spin electrons may be localized.$^{5,12}$
But depending on the doping level, there are some minority spin electrons in 
the extended $t_{2g}\downarrow$ band. Hence they are also paired. 
The resulting p-wave superconducting state will be similar to the ABM 
state.$^{16}$

The apparent finite resistivity may be caused by  the {\sl frustrated} 
p-wave pairing condensation, since the domain wall structures 
of the ferromagnetic phase and the existence of the localized minority spin 
electrons limit the superconducting coherence seriously.
The possible existence of the canted phase$^{12}$ and the electronic phase 
separation$^{17}$ will also give rise to the finite resistivity.
A similar finite resistivity was found in very narrow superconducting metal 
wires where the superconducting condensation is also frustrated.$^{18}$
 
In Section 2 we critically reinvestigate the existing experimental data
of resistivity, specific heat, and tunneling.
Other data will be considered elsewhere. 
Because the superconducting behaviors are mostly screened to some extent,
we need to analyze the existing data with caution. 
The possible p-wave superconducting states are described in Section 3. 
In Section 4 we discuss how to make good room temperature
superconductors for practical applications.

\section* {\bf 2. Reinvestigation of Experimental Data}

It is amazing that the p-wave superconducting condensation 
in manganese oxides was unnoticed until now even though there are so many 
experimental manifestations of the existence of the p-wave 
pairing.$^{1-4,6,7,9,19-24}$ 
It is crucial to notice that the apparent finite resistivity
is caused by some extrinsic reasons not by the intrinsic reasons. 
The possible causes of the nonzero resistivity are the following: 

\noindent
(1). If the ferromagnetic phase forms the domain wall structures, 
the p-wave superconducting phase coherence is established in each
domain and the anisotropic superconducting domains are coupled by the
Josephson coupling.  Then, this leads to the strong cancellation of the 
supercurrents as well as the finite resistivity.

\noindent
(2). The minority spin electrons coming from the both Mn 3d and O 2p
orbitals may hinder the supercurrents and give rise to the
finite resistivity.

\noindent
(3). The homogeneous canted phase developed by de Gennes$^{12}$ or
the coexistence of ferromagnetic (FM) and antiferromagnetic (AF) 
microdomains$^{17}$ will certainly give rise to the finite resistivity.

\noindent
(4). Finally, since supercurrents in $A_{1}$-like state are very anisotropic,
they may not be easily observable as in the case of the A phase of 
superfluid He-3.$^{15}$

\noindent
More conclusive proof of the p-wave pairing in manganese oxides
is the steep jump of the specific heat and the gap opening in 
tunneling.$^{6, 21-24}$ 

\subsection* {\bf 2.1 Resistivity and CMR}

In this Section we consider the resistivity data.$^{19,20}$
Almost all data show the abrupt decrease of the resistivity by
several orders when the temperature is decreasing from the transition 
temperature $T_{c}$.
But it appears that the $T=0$ resistivity does not vanish and is 
$\sim 10 \mu\Omega cm$ for good samples. 
The main obstacle to the realization of the superconductivity in manganese
oxides is the apparent finite resistivity. 
Considering the possibility of {\sl frustrated} p-wave pairing, 
the crucial point is not the finite value of the resistivity 
but the abrupt drop of the resistivity as a function of the temperature.
Notice the similarity between the finite resistivity of
very narrow superconducting metal wires
and that of manganese oxides.$^{18}$
People think that if superconducting and normal electrons are
coexisting, the resistivity will be zero because of the
short-circuiting behavior of the superconducting electrons.
However,  in this case the superconducting coherence may be seriously limited
by the domain wall structures, the effect of the minority 
spin electrons, and some magnetic inhomogeneities.  

Figures 1 (a) and (b) show the temperature dependence of the magnetization and
the resistivity of a single crystal of 
$\rm{La_{0.65}(PbCa)_{0.35}MnO_{3}}$ (LCPMO), respectively.$^{19}$
Data are due to Liu et al..$^{19}$
The transition temperature $T_{c}$ is $300 K$.
It is obvious that the ferromagnetic-to-paramagnetic transition and the 
metal-insulator transition occur simultaneously. The long tail of the 
resistivity seems to be due to the disorder caused by the co-doping of 
Pb and Ca. 
The maximum CMR was observed exactly at $T_{c}$, which indicates the
close relation between superconductivity and CMR. 
In this type of p-wave superconducting state, $T_{c}$ is increasing 
for strong magnetic fields because the ferromagnetic domain size is increasing
by the alignment of the $S=3/2$ core spins. 
Accordingly, the concomitant superconducting fluctuations lead to enhanced 
conductivity and CMR.
Figure 1 (c) also shows the temperature dependence of the resistivity of
$\rm{La_{0.72}Ca_{0.25}MnO_{z}}$ (LCMO) measured by Chahara et al..$^{20}$
It appears that these data also indicate the p-wave superconducting 
condensation. 

\subsection* {2.2 Specific Heat}

The second evidence for the p-wave pairing in manganese oxides
is found in the electronic specific heat measurements.$^{21,22}$ 
The observed steep jump of the electronic specific heat means
the second order phase transition of the electronic state.
But the $T^{3}$ law near $T=0$ was not confirmed clearly.
The reason is certainly related with the apparent finite resistivity.
It appears that the {\sl frustrated} p-wave superconducting electrons and
normal electrons are coexisting. 
Note that the minority spin electrons and holes caused by divalent ion 
doping also contribute to the specific heat near $T=0$.

Figure 2 (a) shows the specific heat data of $\rm{La_{0.8}Ca_{0.2}MnO_{3}}$.
Data are due to Tanaka and Mitsuhashi.$^{21}$
The broken line denotes a lattice heat capacity.
It is clear that the electronic transition is of second order as noted by the
authors, which is consistent with ferromagnetic-metal-to-p-wave superconductor
transition.
Figure 2 (b) shows the electronic specific heat divided by temperature,
$C/T$, for $x=0.33$ $\rm{La_{1-x}Ca_{x}MnO_{3}}$ sample.$^{22}$
Data are from Ramirez et al..$^{22}$ Again we find a steep
jump of the specific heat. Since both the electronic and magnetic 
transitions occur at the same time, the specific heat change also includes 
the magnetic contribution. 
Note that the shape of the specific heat change should be symmetric with 
respect to $T_{c}$ if only the magnetic transition had occurred. 
 
\subsection* {\bf 2.3 Tunneling}

The final evidence for the p-wave pairing superconducting state
is provided by the tunneling data.$^{6,23,24}$
The quasiparticle density of states (DOS) for the ABM or $A_{1}$ state
is parabolic at low energies.
It also shows divergences as the energy is approached to the maximum gap 
energy $\Delta_{o}$ both from below or above. The theoretical DOS will be 
shown in Fig. 5 (b).

Figure 3 (a) shows the normalized conductance versus the sample bias voltage
for $\rm{La_{0.7}Ca_{0.3}MnO_{3}}$ (LCMO). Data are from Wei, Yeh and 
Vasquez.$^{23}$
They observed the density of states of an epitaxial film sample of 
ferromagnetic LCMO by scanning tunneling spectroscopy.
It is noteworthy that this figure is very similar to the DOS of
the ABM  and the $A_{1}$ states.  
The two peaks at $\sim \pm 0.4 V$ imply that the maximum Cooper pair binding
energy equals to $\sim 0.4 eV$. The energy is huge enough to give rise to
the p-wave superconducting state with $T_{c}=260 K$.
The pronounced peaks at $\sim \pm 1.75 V$ may be related with the 
exchange-split, spin-polarized peaks in the density of states
calculated by Pickett and Singh.$^{5}$
In other words, these are half-metallic peaks. 
Figure 3 (b) shows the dynamic conductance $dI/dV$ vs bias $V$ at
$T=4.2 K$ of magnetic tunnel junctions of $\rm{La_{0.67}Sr_{0.33}MnO_{3}}$.
Data are due to Lu et al..$^{6}$
Notice the depletion of the DOS at the zero bias. If we disregard 
some background contribution, this curve seems to show the
zero DOS at the Fermi level, which  agrees with the DOS of the
p-wave pairing state.  
In fact, Hudspeth et al.$^{24}$ clearly observed the depletion in the 
conductance of LCMO films at the Fermi level with decreasing temperature.
The parabolic shape of the conductance curve is also 
consistent with the DOS of the ABM and the $A_{1}$ states.
Since the transition temperature $T_{c}$ is $347 K$, the peaks may
appear at $\sim \pm 0.53 V$ in view of Fig. 3 (a).

\section* {\bf 3. Frustrated P-wave Superconducting State}

Now we discuss the theoretical model for the p-wave superconducting
state which may be realized in manganese oxides.
Since the double exchange model and the related strong interaction 
between spin, charge, and lattice degrees of freedom are not solvable yet,
our discussion is rather qualitative. We don't discuss the mechanism of the
p-wave pairing either. 
Presumably, the strong electron-phonon interaction and the ferromagnetic
interaction lead to the p-wave superconducting condensation in manganese
oxides. 
Nevertheless we can explain the experimental  data of the specific heat and the 
quasiparticle density of states (DOS) in terms of the $A_{1}$ state or the
ABM state, because they are determined entirely by the topology of the zero gap.
In this case even the crystal structure does not play an important role.
On the other hand, if the mobile minority spin electrons exist for some
doping concentrations,$^{5,7}$ the ABM state may be more appropriate.

\subsection*{\bf 3.1 Frustration}

The frustration may be caused by the domain wall structures of the 
ferromagnetic phase and the coexisting minority spin electrons.
Figure 4 (a) shows the domain wall separating domains. 
The spins of the superconducting electrons in the left domain are pointing
upward, whereas those of the electrons in the right domain are
pointing downward.   In the transition region, the conduction electrons may
become normal because the spin directions of the core $t_{2g}$ electrons are 
strongly varying. The two superconducting domains are coupled 
by the Josephson-like coupling. Then, the supercurrents 
decay after passing the transition region and consequently the finite
resistivity is obtained.
Figure 4 (b) shows the effect of the minority spin electrons localized
randomly. The spins of the Cooper pair electrons are canted by the spins of 
the minority electrons.$^{25}$ Then the quasiparticles near the nodes will be
excited and lead to the finite resistivity.
On the other hand, superconductivity is also possible in the canted 
phase.$^{25}$ However, the condensation energy will be much smaller than
that of the ferromagnetic case and the quasiparticles are
easily excited by any perturbation.
Recently, Allodi et al.$^{17}$ reported the possibility of intrinsic
phase separation of the holes into FM microdomains. This may also lead 
to the frustrations and the finite resistivity.

\subsection*{\bf 3.2 $A_{1}$ State and ABM State}

In the ferromagnetic phase the majority spin electrons are free to move, 
whereas minority spin electrons may be localized.
Therefore, only majority (up) spin electrons are paired since
the pairing interaction for the localized electrons will be exponentially
small.$^{26,27}$  
The resulting p-wave superconducting wavefunction is given,
\begin{equation}
{\tilde \phi}_{P} = {\prod'}_{\vec k}(u_{\vec k} + v_{\vec k}
a^{\dagger}_{{\vec k}\uparrow} 
a^{\dagger}_{{-\vec k}\uparrow})|0>,
\end{equation}
where the product over $\vec k$ is restricted to a half-space and
$a^{\dagger}_{{\vec k}\uparrow}$ creates a majority spin electron at  
the $e_{g}$ conduction band.
Because of the Fermi statistics, we have 
\begin{equation} 
u_{-\vec k}=u_{\vec k}, \quad \quad v_{-\vec k}=-v_{\vec k}.
 \end{equation} 
The gap parameter for the $A_{1}$ state is given by 
\begin{equation} 
\Delta_{\vec k}^{\uparrow\uparrow} = \Delta_{o}sin\theta_{\vec k}
e^{i\phi_{\vec k}},
\end{equation}
where $\theta_{\vec k}$ and $\phi_{\vec k}$ are the polar and azimuthal
angles of $\vec k$.
Notice that the gap vanishes at two points along ${\hat k_{z}}={\hat {\bf l}}$
with ${\bf l}$ denoting the angular momentum.
In the presence of the mobile minority spin electrons, the gap parameter of
the ABM state is given by
\begin{equation} 
\Delta_{\vec k}^{\uparrow\uparrow} = \Delta_{o}sin\theta_{\vec k}
e^{i\phi_{\vec k}},\quad
\Delta_{\vec k}^{\downarrow\downarrow} = \Delta_{o}'sin\theta_{\vec k}
e^{i\phi_{\vec k}},
\end{equation}
where $\Delta_{o}'$ is not always the same as $\Delta_{o}$.

Since the specific heat and the density of states for the $A_{1}$ state
and the ABM state show the qualitatively same behaviors, we consider  
those of the ABM state with $\Delta_{o}'=\Delta_{o}$.
The single-particle excitation energy is given by
\begin{equation}
E_{\vec k}=\sqrt{\epsilon_{\vec k}^{2} + \Delta_{o}^{2}sin^{2}
\theta_{\vec k}}.
\end{equation}
Accordingly, the specific heat is found to be
\begin{eqnarray}
C_{V} &\equiv& T{dS\over dT} \nonumber\\
 &=& {1\over 2T^{2}}\sum_{\vec k}[E^{2}_{\vec k} - {T\over 2}
{d \Delta_{\vec k}^{2} \over dT}]sech^{2}{E_{\vec k}\over 2T},
\end{eqnarray}
where S denotes the entropy. 
In the low temperature limit, one finds
\begin{equation}
C_{V} = 3.5(\gamma T_{c})({T\over T_{c}})^{3},
\end{equation}
where $\gamma = C_{n}/T$. $C_{n}$ is the normal specific heat.
The specific heat jump for the ABM state is given by
\begin{equation}
{\Delta C_{V}\over C_{n}} \cong 1.19.
\end{equation}
(For the $A_{1}$ state the jump is 0.74.$^{15}$)
The normalized single-particle density of states (DOS), $N(E)/N_{o}$, 
was found to be$^{16}$
\begin{equation}
{N(E)\over N_{o}} = {|E|\over 2\Delta_{o}}
ln\left|{E+\Delta_{o}\over E-\Delta_{o}}\right|,
\end{equation}
where $N_{o}$ is the normal DOS. 
The $N(E)$ is parabolic, that is, $N_{o}(E/\Delta_{o})^{2}$,
near the Fermi energy and becomes infinite at $E=\Delta_{o}$.     

Figures 5 (a) and (b) show the specific heat and the normalized density of
states, $N(E)/N_{o}$, for the ABM state with the transition temperature 
$T_{c}=100K$. 
Comparing with Figs. (2) and (3), the agreements between
our theoretical curves and experimental data are excellent. 
It is reasonable to conclude that the p-wave pairing superconducting 
condensation is realized in manganese oxides. 
However, the superconducting behaviors are almost screened by 
domain wall structures, the existence of the localized minority spin electrons,
the canted phase, and the coexistence of the FM and AF microdomains. 

\subsection* {\bf 3.3 $A_{1}$-like State in cubic crystalline symmetry}

In manganese oxides, the Mn ions form a cubic lattice.
The group theoretical classification of p-wave pairing superconducting states
under cubic symmetry has been already done.$^{28-32}$ 
The superconducting classes with ferromagnetic properties are
obtained from three dimensional representations of the
cubic group. They are $D_{3}(E) $ and $D_{4}(E)$ classes from
three-dimensional representations $F_{1}$ and $F_{2}$.$^{29}$
The phases $D_{3}(E)$ have zeros at the points of intersection
of the Fermi surface with a three-fold axis, whereas the phases 
$D_{4}(E)$ have zeros at the points of intersection of the Fermi surface
with a four-fold axis.  
$D_{3}(E)$ and $D_{4}(E)$ phases are eight-fold and six-fold degenerate, 
respectively.  These superconducting states resemble 
the A phase of He-3.  
On the other hand, manganese oxides seems to be coverted to
a rhombohedral structure with the Jahn-Teller effect.$^{33}$ 
The corresponding p-wave superconducting classes have also been 
investigated.$^{34}$ 
>From these states we may obtain proper $A_{1}$-like and ABM-like 
p-wave pairing superconducting states which may be realized in manganese 
oxides.  More details will be reported elsewhere.

\section* {\bf 4. Discussion}

More studies are required to check whether the {\sl frustrated} p-wave 
pairing superconducting condensation is really realized in manganese oxides.
When analyzing the experimental data, it is important to take into account
the screening effect caused by domain wall structures, minority spin
electrons, and other magnetic inhomogeneities.
The magnetic properties of manganese oxides need to be reinvestigated.
In particular, NMR experiments probing the spin dynamics may provide the 
specific information on the structure of the p-wave superconducting state.
The collective modes such as sound modes and spin-wave modes
may also give a clue to the existence of the p-wave pairing.
The effects of magnetic fields are particularly interesting, since
they give rise to CMR and metal-insulator transition in some samples.$^{35}$
The spin-polarized tunneling experiment$^{6}$ should be done up to 
higher bias voltages to find the gap opening.

Let us discuss how we can make good room temperature superconductors.
It depends on how we eliminate the frustrations and make a good single crystal 
with one single magnetic domain and 100 \% spin polarization near the 
Fermi level.
It is also desirable for the minority spin electrons to have a large 
insulating gap which results in the complete disappearance of the minority 
spin DOS near the Fermi surface.
In other words, we need a perfect half-metallic ferromagnet. 
The local spin distortions caused by any possible bound or self-trapped
states of the carriers should be eliminated since they give rise to
the canting of the Cooper pair spins and limit the superconducting
coherence seriously.
If we overcome the above problems, it is highly likely that we can make 
almost room temperature superconducting manganese oxides which
have the zero temperature resistivity smaller than that of the good Cu samples 
by a factor of more than a thousand.
 
\section* {5. Conclusion}
We have proposed the {\sl frustrated} p-wave pairing superconducting
state to explain CMR, the sharp drop of resistivity, the steep jump of
specific heat, and the gap opening in tunneling of manganese oxides.
The p-wave pairing state is similar to the $A_{1}$ state in superfluid
He-3. The apparent finite resistivity may be caused by the domain wall
structures of the ferromagnetic phase, the coexisting minority spin
electrons, the canted phase,  and the possible existence of the
FM and AF microdomains. If we eliminate the above
problems, the resistivity of manganese oxides will be smaller than that of
good Cu samples by a factor of more than a thousand.
This study opens up the possibility of room temperature superconductivity.

\vspace{2pc}
\centerline{\bf Acknowledgments}

This work has been supported by the Brainpool project of KOSEF and the MOST.
I am grateful to Dr. H. T. Kim and Profs. Jong-Jean Kim, D. Youm, K. J. 
Chang, B. I. Min, and K. T. Park for discussions.

\begin{figure}
\caption{(a) The temperature dependence of the magnetization of a single 
crystal of LPCMO, taken from Liu et al. (Ref. 19). The Curie temperature
$T_{c}$ is  $300 K$.
(b) The resistivity vs temperature in zero field and in a 5.5-T field,
taken from Liu et al. (Ref. 19).
(c) The temperature dependence of the resistivity at zero field, taken
from Chahara et al. (Ref. 20).}
\end{figure}

\begin{figure}
\caption{ (a) The temperature dependence of specific heat of
monoclinic $\rm{La_{0.8}Ca_{0.2}MnO_{3}}$. (After Tanaka and
Mitsuhasi Ref. 21.) 
The broken line denotes a lattice specific heat calculated by a Debye model.
(b) specific heat divided by temperature, C/T for $x=0.33$ $\rm{La_{1-x}
Ca_{x}MnO_{3}}$ with lattice contribution subtracted. 
(After Ramirez et al. Ref. 22.)}
\end{figure}

\begin{figure}
\caption{ (a) STM spectroscopy data at 77K, plotted as the normalized 
conductance $(dI/dV)/(\overline{I/V})$ vs the bias voltage V, 
taken from Wei et al. (Ref. 23).
(b) dynamic conductance $dI/dV$ vs bias V at $T=4.2 K$ of a magnetic
tunnel junction, taken from Lu et al. (Ref. 6).}
\end{figure}

\begin{figure}
\caption{ (a) The triplet Cooper-pairs in the domain wall
structures of the ferromagnetic phase. (b) a Cooper pair in the 
presence of the minority spin electrons localized randomly.}
\end{figure}

\begin{figure}
\caption{ (a) Specific heat vs temperature for the ABM state.
The transition temperature $T_{c}$ is 100K.
(b) the density of states vs energy divided by the maximum gap parameter,
$E/\Delta_{o}$, for the ABM state.}
\end{figure}
\end{document}